\documentclass[
reprint,
superscriptaddress,
amsmath,amssymb,
aps,
prb,
twocolumn
]{revtex4-2}
\usepackage{tikz}
\usetikzlibrary{arrows.meta, positioning, shapes.geometric}
\usepackage{graphicx}
\usepackage{dcolumn}
\usepackage{bm}
\usepackage[breaklinks]{hyperref}
\hypersetup{colorlinks=true, linkcolor=blue, citecolor=blue, filecolor=blue, urlcolor=blue}
\graphicspath{{figures}}
\usepackage{bm}

\usepackage{graphicx,subfigure}
\usepackage{epsfig}
\usepackage{bm}
\usepackage{dcolumn}
\usepackage{color}
\usepackage{physics}
\usepackage{float}
\makeatletter
\newcommand*{\rom}[1]{\expandafter\@slowromancap\romannumeral #1@}
\makeatother
\usepackage{lipsum}
\usepackage{subfigure}
\usepackage{dsfont}
\usepackage{enumitem}
\usepackage{tikz}
\usetikzlibrary{quantikz}
\usepackage{amsmath} % Required for \boldsymbol

\usepackage[normalem]{ulem}
\newcommand{\rz}[1]{\textcolor[rgb]{0,0.0,0}{#1}}
\newcommand{\hzc}[1]{\textcolor[rgb]{0,0.0,0.0}{#1}}

\begin{document}

\title{Benchmarking Quantum Solvers in Noisy Digital Simulations for Financial Portfolio Optimization} 
\author{Ruizhe Shen}
\email{e0554228@u.nus.edu}
\affiliation{Department of Physics, National University of Singapore, Singapore 117542}
\author{Zichang Hao}
\affiliation{Department of Physics, National University of Singapore, Singapore 117542}
\author{Ching Hua Lee}
\email{phylch@nus.edu.sg}
\affiliation{Department of Physics, National University of Singapore, Singapore 117542}
\date{\today}
\begin{abstract}  
In this work, we benchmark two prominent quantum algorithms: Quantum Imaginary-Time Evolution (QITE) and the Quantum Approximate Optimization Algorithm (QAOA) for obtaining the ground state of Ising-type Hamiltonians. Specifically, we apply them to the Markowitz portfolio optimization problem in quantitative finance, on both digital quantum computers and local quantum simulators with controllable two-qubit errors (noise).
In noiseless settings, we find that QAOA achieves excellent convergence to the optimal results.  Under noisy conditions, the QITE method exhibits greater robustness and stability, though it incurs substantially more classical numerical cost.  In contrast, we demonstrate that QAOA offers better scalability and can still yield robust results if the noise can be effectively mitigated. Our findings provide valuable insights into the trade-offs between scalability and noise tolerance and demonstrate the practical potential of quantum algorithms for solving real-world optimization problems on near-term quantum devices.
\end{abstract}
\maketitle

\section{Introduction}\label{sec0}
Leveraging on quantum computing for real-world applications represents a promising yet challenging frontier \cite{preskill2018quantum, pelofske2022quantum,lau2022nisq,torlai2020machine,fauseweh2024quantum,kim2023evidence,zhang2021implementation, ezratty2023we}. With its ability to process and manipulate
information with qualitatively heightened efficiency, quantum processors have significant potential to outperform classical resources in solving complex problems across domains from physics \cite{daley2022practical,brown2010using,martinez2016real,paulson2021simulating,zhu2021probing,chen2022error,frey2022realization,fauseweh2024quantum,smith2019simulating,koh2022simulation,koh2022stabilizing,chen2023high,chen2023robust,shen2025circuit,shen2025robust,shen2025observation,koh2024realization,shen2024enhanced,chen2024direct,koh2025interacting,desaules2024robust} to practical applications such as cryptography and drug discovery  \cite{mavroeidis2018impact, fernandez2020towards, bernstein2017post,pirandola2020advances,cao2018potential,blunt2022perspective}.  More recently, quantum computing has demonstrated substantial promise in tackling \rz{computationally demanding} optimization problems \cite{farhi2014quantum,zhou2020quantum,ajagekar2019quantum,au2023np, mouton2024deep,chen2025benchmarking}.  A particularly relevant example with practical importance is portfolio optimization in quantitative finance  \cite{kerenidis2019quantum,rebentrost2018quantum,grant2021benchmarking,hegade2022portfolio}. This task involves constructing a portfolio that maximizes expected returns, a problem traditionally addressed using classical optimization algorithms and machine learning techniques \cite{ban2018machine,grant2021benchmarking,hegade2022portfolio,gunjan2023brief}.  As the complexity and dimensionality of modern financial portfolios increase, quantum hardware is emerging as a compelling tool for managing sophisticated optimization models.

To solve such optimization problems on digital quantum computers, one first needs to formulate them within a quantum Hamiltonian framework, such that the optimal solution corresponds to the ground state of a spin-1/2 system. A widely used method is Quantum Approximate Optimization Algorithm (QAOA) \cite{brandhofer2022benchmarking,farhi2014quantum,blekos2024review,zhou2020quantum,willsch2020benchmarking,hadfield2019quantum}. \rz{This approach is implemented through optimizing a parameterized circuit and minimizing the expectation value of a cost function.} Beyond these methods, Quantum Imaginary-Time Evolution (QITE) provides a deterministic route to ground-state preparation by evolving a quantum state along imaginary time \cite{motta2020determining,mcardle2019variational,nishi2021implementation}.

While previous works have explored quantum solvers for portfolio optimization problems \cite{brandhofer2022benchmarking,rebentrost2024quantum,hegade2022portfolio,buonaiuto2023best}, most remain constrained by limitations inherent to current quantum hardware \cite{cai2023quantum,suzuki2022quantum,larose2022error,hassija2020present,rietsche2022quantum}, and few explicitly examine the impact of noise. To assess the practical performance of these methods in noisy simulations, we conduct a comprehensive benchmarking study of these quantum algorithms applied to the Markowitz portfolio optimization problem.  Our results show that QAOA performs exceptionally well under ideal, noiseless conditions; however, it exhibits poor robustness in the presence of realistic noise \cite{resch2021benchmarking,knill2005quantum,mccaskey2019quantum}.  In contrast, QITE exhibits greater robustness to noise, but requires extensive training of parameterized circuits to accurately approximate the desired imaginary-time evolution. Overall, our findings suggest that current quantum optimization strategies are still fundamentally limited by challenges related to noise resilience and scalability. While both methods face limitations on noisy devices, QAOA shows strong potential for large-scale applications if hardware noise can be significantly mitigated through advanced error mitigation techniques or future fault-tolerant architectures \cite{cai2023quantum,endo2018practical,temme2017error,takagi2022fundamental,strikis2021learning,qin2022overview}.

This work is organized as follows: In Section II, we introduce the problem of financial portfolio optimization and the potential advantages of tackling it through a quantum algorithm. Following this, in Section III, we introduce the Markowitz portfolio optimization problem and detail its mapping to a quantum Hamiltonian. In Section IV, we describe the quantum optimization algorithms employed in this study, including QAOA and QITE. Then, in Section V, we present simulation results and compare the performance of QAOA and QITE under both noiseless and noisy conditions. Finally, we summarize our benchmarking results of the two quantum algorithms and discuss applications for future research in Section VI.

\begin{figure*}
    \centering
    \includegraphics[width=0.9\linewidth]{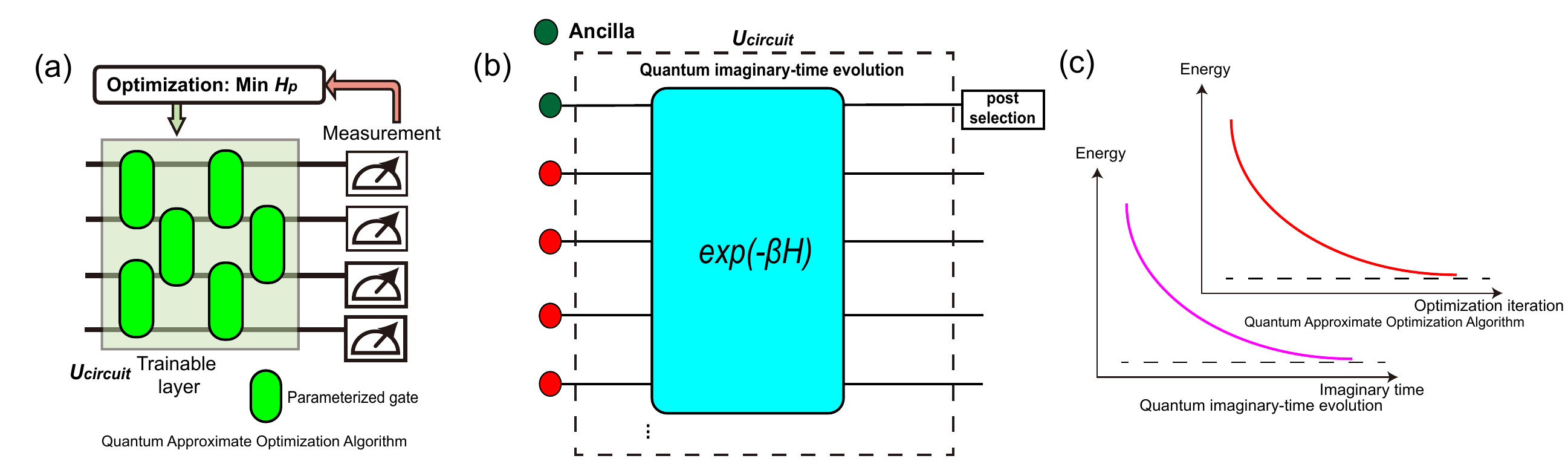}
    \caption{Schematics of the Quantum Approximate Optimization Algorithm (QAOA) and Quantum Imaginary-Time Evolution (QITE) approaches. (a) The QAOA framework, where a parameterized quantum circuit is iteratively optimized to minimize the expectation value of the target Hamiltonian $H_p$. (b) Quantum circuit structure for implementing QITE.
 The green qubit represents the ancilla qubit, and the red qubits correspond to the system qubits used for encoding the problem instance.   The system qubits are coupled to the ancilla qubit to implement the QITE dynamics, $e^{-\beta H}$. Here, $H$ represents the Hamiltonian defined in Eq.~\ref{ising}, whose ground state is to be found. The final result is obtained by post-selecting the ancilla qubit in the $\ket{0}$ state. Additional implementation details are provided in Appendix.~\ref{QITE}. (c) Evolution of QITE and QAOA. The energy of both systems decreases under imaginary time (violet curve) or optimization iterations (red curve) to approximate ground states (dashed line).
}
    \label{fig:circuit}
\end{figure*}

\section{Portfolio optimization problem}
In the financial sector, there exist a variety of optimization problems \cite{tutuncu2003optimization,zenios1993financial,gilli2019numerical,orus2019quantum}. These include forecasting tasks such as pricing, risk assessment, anomaly detection in transactions, and identifying customer preferences, as well as optimization tasks like selecting optimal portfolios, developing effective trading strategies, and designing robust hedging techniques.  Portfolio optimization specifically refers to the problem of identifying the optimal allocation of a given set of assets \cite{hegade2022portfolio,rosenberg2015solving,grant2021benchmarking}. In this section, we provide an introduction to the problem of Markowitz portfolio selection.

\subsection{Markowitz portfolio selection}
Here, we first introduce the mathematical model for Markowitz portfolio selection \cite{cohen1967empirical,rubinstein2002markowitz,zhang2018portfolio,pogue1970extension}.  In this framework, a fixed budget $b$ can be allocated among $m$ assets. Each asset $u$ is characterized by its expected return $r_u$. The model also incorporates a covariance matrix $c_{u,v}$, which quantifies the historical price correlation between assets $u$ and $v$. Here, the task of Markowitz portfolio selection is to construct an optimal portfolio that maximizes expected return while minimizing overall portfolio variance. This formulation is based on Markowitz portfolio theory, which aims to achieve an efficient trade-off between risk and return \cite{grant2021benchmarking}.

Next, we introduce the quantized minimum investment fraction $p_w = 1/ 2^{w-1}$, where $w \in \mathbb{N}$  parametrizes the number of budget slices $2^{w-1}$. The investment fraction for the $u$-th asset can be described as an integer multiple $z_u p_w$, where the integer $z_u$ indicates how many quantized units are allocated to asset $u$. We then impose the constraint that the total investment must match the full budget: $\sum_u z_u p_w =1$. Within this framework, the Markowitz portfolio selection problem is reformulated as an optimization task:  identifying the optimal set of $z_u$ values that maximizes the following objective function \cite{grant2021benchmarking}
\begin{equation}\label{opt}
F(z)=\theta_1 \sum_{u=1}^m r_uz_u -\theta_2 \sum_{u,v=1}^m c_{u,v} z_{u}z_{v}- \theta_3\sum_{u=1}^m (z_u p_w b -b)^2.
\end{equation}
\hzc{Here, $b$ represents the real amount of budget, while the product $z_u p_w$ ($0\leq z_u p_w\leq 1$) denotes the fraction of the budget allocated to asset $u$.}
The parameters $\theta_{1,2,3}$ represent the investor's preferences with respect to expected return, portfolio risk, and tolerance for budget deviation, respectively. Specifically, the $\theta_1$ term rewards high expected return, the $\theta_2$ term penalizes risk arising from asset correlations, and the $\theta_3$ term imposes a penalty for deviation from the total budget \cite{grant2021benchmarking}. \hzc{The inclusion of the factor $b$ in the $\theta_3$ term aligns it with the budget dependence of $r_u$ [Eq.~\ref{eq:r_u}] and $c_{u,v}$ [Eq.~\ref{eq:c_uv}] in the first two terms. This ensures that the budget deviation penalty remains comparable in magnitude to the return and risk terms, even when $b$ is large.}. In our implementations,  we set $\theta_{1}=0.8$, $\theta_{2}=0.1$, $\theta_{3}=0.1$, and $b=10$. More details are presented in Appendix.~\ref{data}.

\subsection{Correspondence between Financial and Quantum Models}
To solve this optimization problem described by Eq.~\ref{opt} on a quantum platform, we first map it onto the ground-state problem of a corresponding quantum Hamiltonian. To achieve this, we rescale the expected return based on historical price data and define
\begin{equation}
	r_u = \frac{b p_w}{a_{u,N_f}}\cdot \sum_{l=1}^{N_f} \frac{a_{u,l+1}-a_{u,l}}{N_f-1}
	\label{eq:r_u},
\end{equation} where $a_{u,l}$ is the $l$-th history of price data for the $u$-th asset and $N_f$ denotes the number of historical assets price points \cite{huang2012mean, martin2017expected}. {In this work, these historical prices are randomly generated, and details are provided in Appendix.~\ref{data}.} The corresponding covariance matrix $c_{u,v}$ is computed as
\begin{equation}
	c_{u,v} = \frac{b^2p_w^2}{a_{u,N_f}a_{v,N_f}} \cdot \frac{ \sum_{l=1}^{N_f} (a_{u,l}-\bar{a}_u) (a_{v,l}-\bar{a}_v)}{N_f -1}.
	\label{eq:c_uv}
\end{equation}
where $\bar{a}$ denotes the mean price of assets.
To enable encoding these data on a qubit system, we represent each integer $z_u$ in binary form:
\begin{equation}
	z_u = \sum_{k=1}^w 2^{k-1} x_{i(u,k)},
\end{equation}
where $i(u,k) = (u-1)w+k$ and $x_i \in \{0,1\}$ is a binary variable indicating the presence or absence of investment in a specific quantized unit.

{For instance, the optimal bitstring shown in FIG.~\ref{fig:ibmqaoa} is $010100100$ ($101011011$ under little-endian qubit ordering). This solution encodes an investment configuration for a portfolio of three assets with three binary slices per asset.   The actual investments are $z_{0}/p_{w}=1/2$, $z_{1}/p_{w}=1/4$, and $z_{2}/p_{w}=1/4$. }

Under this representation, the original optimization problem is reduced to finding the ground state of an Ising Hamiltonian:
\begin{equation}\label{ising}
	H= \sum_i h_i \sigma_i^z + \sum_{i,j} J_{i,j}\sigma_i^z \sigma_j^z + \delta
\end{equation}
where classical binary quantities are mapped to quantum operators via the transformation $x_i \rightarrow \frac{1}{2} (1+\sigma_i^z)$, with the Pauli operator $\sigma^{z}_i$. The coefficients $h_{i}$, $J_{i,j}$ and  $ \delta$ canbe derived from the parameters appearing in Eq.~\ref{opt}, which are computed from the asset prices, as detailed in Appendix~\ref{ap1}.

There exists previous works that demonstrated how this portfolio optimization problem can be addressed using quantum methods \cite{farhi2000quantum,herman2023quantum,grant2021benchmarking,cohen2020portfolio,brandhofer2022benchmarking,rebentrost2024quantum,hegade2022portfolio,buonaiuto2023best,chen2025benchmarking}. However, most focus on noiseless settings, which do not accurately reflect the limitations of noise in current platforms, including gate errors, decoherence, and imperfect measurements. Given these, we implement and benchmark different quantum algorithms %strategies 
on noisy simulators and demonstrate whether these methods can identify optimal results under certain controlled noise levels. In the following section, we will provide a comprehensive discussion of these quantum algorithms and their effectiveness in optimizing financial portfolios.

\section{Methods}
In this section, we outline the methods employed to solve the optimization problem described above using digital quantum processors and their simulators.
\begin{figure}
	\centering
	\includegraphics[width=0.9\linewidth]{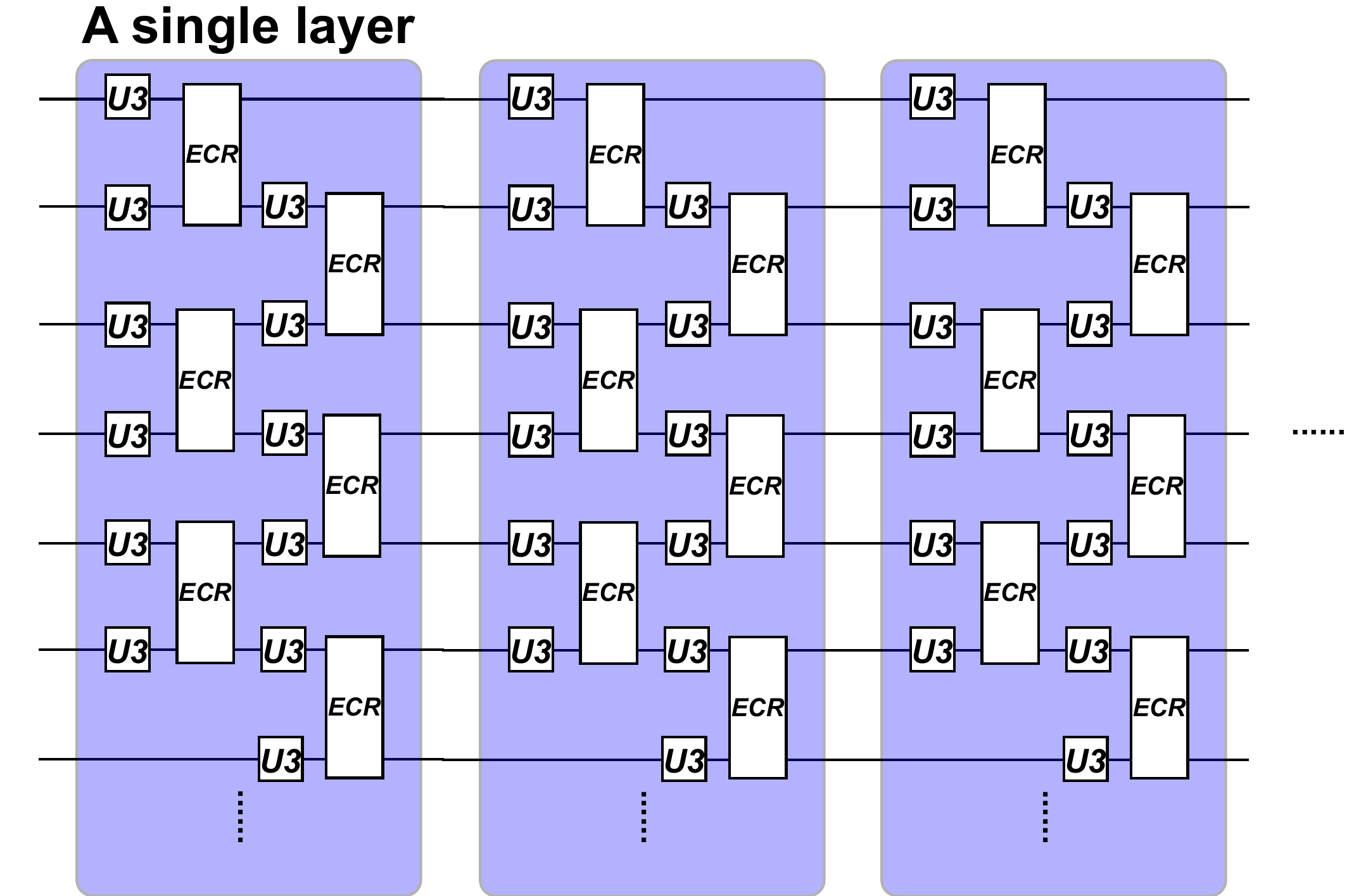}
	\caption{Structure of our trainable variational quantum circuits.  Each layer (purple block) consists of parameterized single-qubit $U_{3}$ gates followed by ECR (Echoed Cross Resonance) gates that entanglement two adjacent qubits. The optimization process is achieved through optimizing the angular parameters in each $U_{3}(\theta, \phi, \lambda)$ gate. For our local noisy simulation, each ECR gate is implemented through a CX gate. The entire circuit in this work consists of $2$ to $8$ of such variational layers, depending on the degree of optimization desired. This structure is used for the implementation of both QITE and QAOA methods.}
	\label{fig:circuit2}
\end{figure}

\subsection{Quantum optimization algorithms}
The Quantum Approximate Optimization Algorithm (QAOA) is a variational algorithm that can be employed to approximate the ground state of a given Hamiltonian \cite{farhi2000quantum,farhi2014quantum}. This method operates by minimizing the expectation of the target Hamiltonian $H_{p}$ through a parameterized quantum circuit. Generally, this is achieved by minimizing the cost function
\begin{equation}\label{ct}
C^{0}(\boldsymbol{s})= \langle \Psi(\boldsymbol{s}) |H_p |\Psi(\boldsymbol{s})\rangle,
\end{equation}
but we later introduce a modification to this form in order to accelerate convergence. Here, $\boldsymbol{s}$ is a generalized vector containing the entire set of variational parameters $(s_{0}, s_{1},...)$, where each $s_{i}$ denotes the angles in each rotational gate, as defined in Eq.~\ref{u3}. These variational angles are optimized by minimizing the difference between the exact ground-state energy $E_{g}$ and the estimated energy $E(\boldsymbol{s})$ i.e. by minimizing the cost function
\begin{equation}\label{cost}
C(\boldsymbol{s})=|E_{g}- E(\boldsymbol{s})|.
\end{equation} 
This setup is schematically shown in FIG.~\ref{fig:circuit} (a). \rz{Note that $E_{g}$ is the exact ground level, which is a relative energy shift. In general, this term is not strictly necessary, since QAOA can be implemented by directly minimizing the cost function defined in Eq.~\ref{ct}. However, incorporating the relative shift often accelerates convergence, thereby reducing the number of iterations and lowering the overall simulation cost.} 

\rz{To simulate this optimization process on quantum circuits, we first evaluate the expectation value $E(\boldsymbol{s})$, obtained by measuring the expectation of each local term in the Hamiltonian. We then iteratively adjust the parameters in $\boldsymbol{s}$ to minimize the deviation $|E_{g}- E(\boldsymbol{s})|$. After each update, the new parameters are re-uploaded to the quantum circuit, and the cycle of measurement and optimization is repeated until the cost function reaches convergence within a reasonable tolerance. For the classical optimization step, we primarily employ the COBYLA (Constrained Optimization BY Linear Approximations) optimizer to achieve optimization \cite{polak2012optimization}. As an alternative, we also test the L-BFGS-B (Limited-memory Broyden–Fletcher–Goldfarb–Shanno with Box constraints) algorithm \cite{polak2012optimization}. IWhile both methods are viable, we consistently observe that COBYLA achieves faster and more reliable convergence in practice. }

Through an alternating evolution process, QAOA efficiently explores the Hilbert space to approximate the ground state. The specific structure of the variational circuit used in this work is detailed in section~\ref{vqe}. While QAOA has demonstrated strong effectiveness and versatility across a variety of optimization tasks, its performance is not robust in the presence of noise in tasks that require high-fidelity identification of optimal quantum states, as we will demonstrate below. In the following section, we conduct a detailed evaluation of QAOA performance in solving this problem, with a particular focus on its robustness in noisy environments and its effectiveness in approximating optimal quantum states.

\begin{figure*} 
	\centering
	\includegraphics[width=0.85\linewidth]{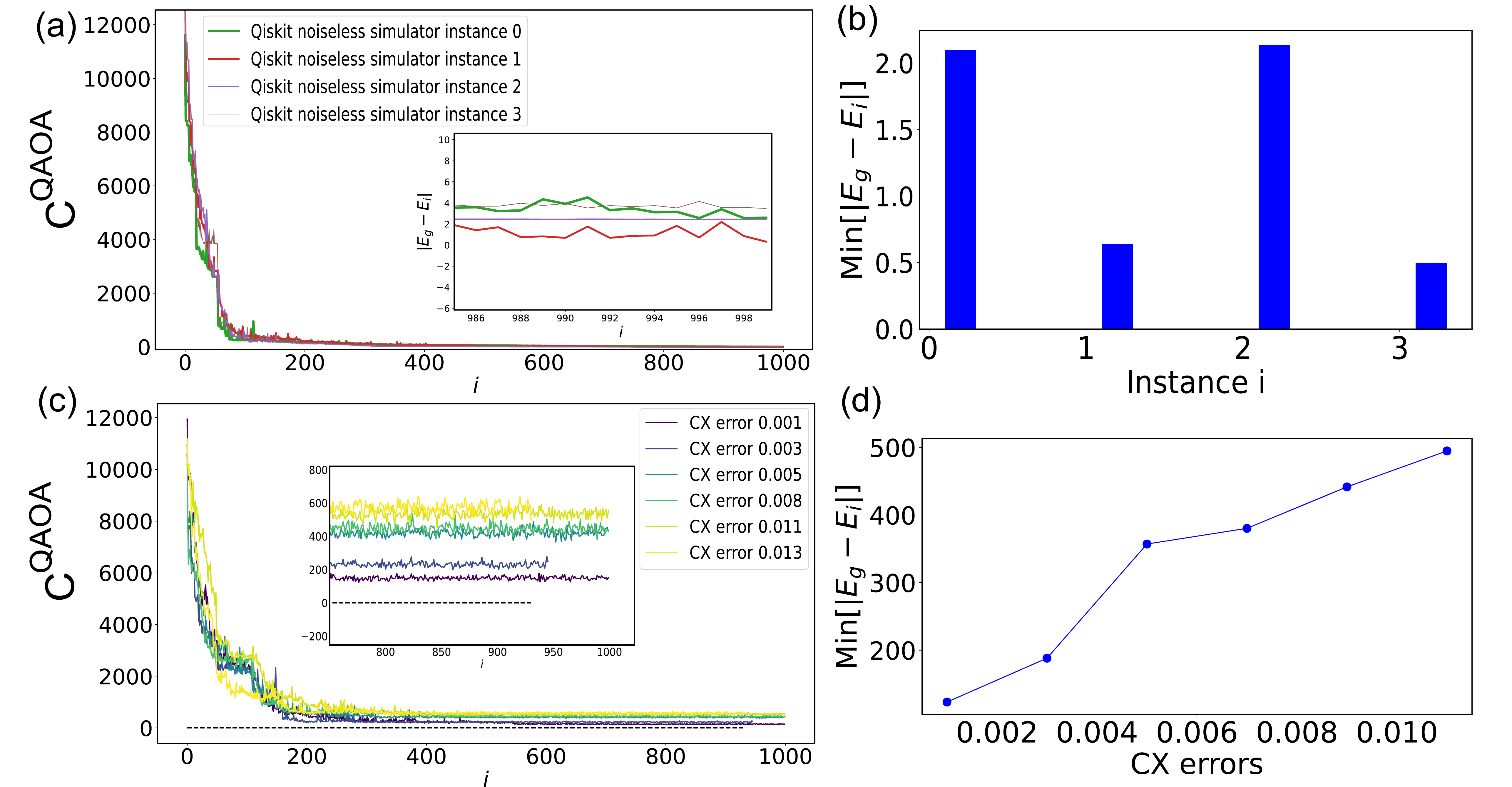}
	\caption{Evaluation of QAOA robustness in estimating the ground-state energy under noiseless and noisy simulations. Here, the model is defined by Eq.~\ref{ising}, $i$ denotes the number of optimization iterations, and the cost function [Eq.~\ref{cost}] quantifies the deviation between the exact ground-state energy and the estimated energy during the optimization process.  
	 (a)  Convergence of the cost function for four representative problem instances simulated on a noiseless Qiskit backend. All instances demonstrate rapid and consistent convergence. The inset zooms in onto the deviation of $E_i$ from the true ground state energy $E_g$ at large $i$, with small fluctuations indicative of good convergence. (b)  Minimum energy deviations (minimum values of cost function throughout optimization) for each instance from (a), all exhibiting the good convergence $C^{\prime\prime}<2.5$. (c) Convergence behavior of the QAOA simulations on local noisy simulators under different CX errors rates. As noise increases, the optimization becomes unstable and energy estimates deviate significantly from the ground state, as evidenced by the significant oscillations shown in the inset. (d) Minimum energy deviations as a function of CX error levels, where each point corresponds to the lowest $|E_{g}-E_i|$ in (c). As the CX error rates increase, the ground state convergence also deteriorates roughly proportionally. For (c) and (d), we select the first instance in (a) here. \rz{Note that different instances correspond to Hamiltonians with distinct parameters corresponding to different sets of generated asset data, and more details are shown in Appendix.~\ref{data}.} The system size for the Hamiltonian is $L=9$ {qubits}.}
	\label{fig:qaoa}
\end{figure*}

\subsection{Quantum imaginary-time evolution}
We next discuss the Quantum Imaginary Time Evolution (QITE) method. This approach obtains the ground state of a quantum system via the following evolution:
\begin{equation}\label{qite}
	\ket{\psi}_{\bf ground}\approx e^{-\beta H}	\ket{\psi(0)}/\sqrt{||e^{-\beta H}	\ket{\psi(0)}||},
\end{equation}
where $H$ is the system Hamiltonian, and $\beta$ is the imaginary time parameter \cite{motta2020determining,mcardle2019variational}. As $\beta$ increases, excited states are suppressed, and the evolution converges towards the ground state of the system. 
Since this process is intrinsically non-unitary, its implementation on quantum hardware requires an effective embedding strategy: the non-unitary operator $e^{-\beta H}$ is realized through an extended unitary operation  which we denote as $U_\text{circuit}$, as schematically illustrated in FIG.~\ref{fig:circuit} (b). In this setup, an ancilla qubit (shown in green) is coupled to the physical qubits to implement the QITE process. Then, post-selection on an ancilla qubit enables the extraction of the desired state $e^{-\beta H}	\ket{\psi(0)}/\sqrt{||e^{-\beta H}\ket{\psi(0)}||}$. Further technical details are provided in Appendix~\ref{QITE}.

\subsection{Variational circuits for implementing QAOA and QITE}\label{vqe}

To implement QAOA and QITE on an actual quantum circuit, we employ variational circuits i.e. parametrized quantum circuits to be optimized. The structure of a variational circuit $V$ is shown in FIG.~\ref{fig:circuit2}, where the variational parameters enter each of the $U_{3}$ gates
\begin{equation}\label{u3}
	U_{3}(\theta, \phi, \lambda)=\left[\begin{array}{cc}
		\cos \left(\frac{\theta}{2}\right) & -e^{i \lambda} \sin \left(\frac{\theta}{2}\right) \\
		e^{i \phi} \sin \left(\frac{\theta}{2}\right) & e^{i(\phi+\lambda)} \cos \left(\frac{\theta}{2}\right)
	\end{array}\right],
\end{equation}
which is parameterized by three tunable angles $\theta, \phi, \lambda$. 
The entire variational circuit $V$ is thus parameterized through the collection of all the vectors $s_{i}=(\theta_{i}, \phi_{i}, \lambda_{i})$.

Variational circuits are used differently in QITE and QAOA.
Exactly representing the target unitary $U_{\rm circuit}$ used in the QITE process (corresponding to the circuit shown in FIG.~\ref{fig:circuit} (b)) as a quantum circuit is typically infeasible, as it may involve circuit depths that are orders of magnitude greater than what is allowed by current NISQ noise thresholds. However, a properly optimized variational circuit $V$ can approximate $U_{\rm circuit}$ rather accurately through the minimization of the following cost function% to achieve the convergence $V\approx U_{\rm circuit}$
\begin{equation}\label{E}
\begin{aligned}
C^{{\text{QITE}}}(\boldsymbol{s})=1-{\rm Tr}[V^{\dagger}(\boldsymbol{s})U_{\rm circuit}]/2^{n}\\
=1-\bra{\psi_{h}}V^{\dagger}(\boldsymbol{s})U_{\rm circuit}\ket{\psi_{h}}
\end{aligned}	
\end{equation}
where $n$ is the number of qubits in the entire circuit, and $\ket{\psi_{h}}=\text{Hadamard}^{\otimes n}\ket{000...}$. In the well-optimized limit where $C^{{\text{QITE}}}(\boldsymbol{s})\approx 0$, we have $V\approx U_{\rm circuit}$.

For the QAOA algorithm, %the cost function is reformulated as: CH: misleading to say it is reformulated. 
the variational circuit $V$ is optimized to minimize the discrepancy between the energy expectation and the ground state energy, instead of maximizing its overlap with a desired evolution operator.
We write the QAOA cost function Eq.~\ref{cost} as 
\begin{equation}\label{cost2}
C^{{\text{QAOA}}}(\boldsymbol{s})=|E_{g}-\bra{\psi(\boldsymbol{s})}H\ket{\psi(\boldsymbol{s})}|,
\end{equation}
where the expectation of the Hamiltonian $H$ is taken with respect to ${\ket{\psi(\boldsymbol{s})}=}V(\boldsymbol{s}))\ket{\psi_{0}}$, the state evolved from $\ket{\psi_{0}}=\ket{000...}$ under the variationally optimized operator $V$.
%where we have calculated the expectation of the Hamiltonian $H$ under the evolved state $\CH{\ket{\psi(\boldsymbol{s})}=}V(\boldsymbol{s}))\ket{\psi_{0}}$, with the initial state being $\ket{\psi_{0}}=\ket{000...}$. 

From FIG.~\ref{fig:circuit2}, the variational circuit also consists of (non-parametrized) two-qubit entangling gates known as Echoed Cross Resonance (ECR) gates, which are natively supported on the IBM Quantum processors that we use. Functionally equivalent  to CX gates, they are defined by
\begin{equation}
E C R=\frac{1}{\sqrt{2}}\left(\begin{array}{cccc}
0 & 1 & 0 & i \\
1 & 0 & -i & 0 \\
0 & i & 0 & 1 \\
-i & 0 & 1 & 0
\end{array}\right).
\end{equation}
However, in our local noisy simulations, it is more straightforward to include noise effects in CX gates through imperfect X operations. Therefore, we use CX gates for all local simulations in this work, with each ECR gate decomposed as $(H\otimes I) CX (I \otimes S^{\dagger}) CX (H\otimes X) $, with $S^{\dagger}$ denoting the S-adjoint gate.

\rz{In our study, the implementation of QITE requires deeper circuits with a larger number of trainable parameters. The 9-qubit instance shown later requires only 2 layers in the QAOA circuit, whereas the corresponding QITE circuit demands between 4 and 8 layers to achieve convergence. Thus,  QITE  in turn results in a substantially higher classical optimization cost} 
\begin{figure*}
    \centering
    \includegraphics[width=0.9\linewidth]{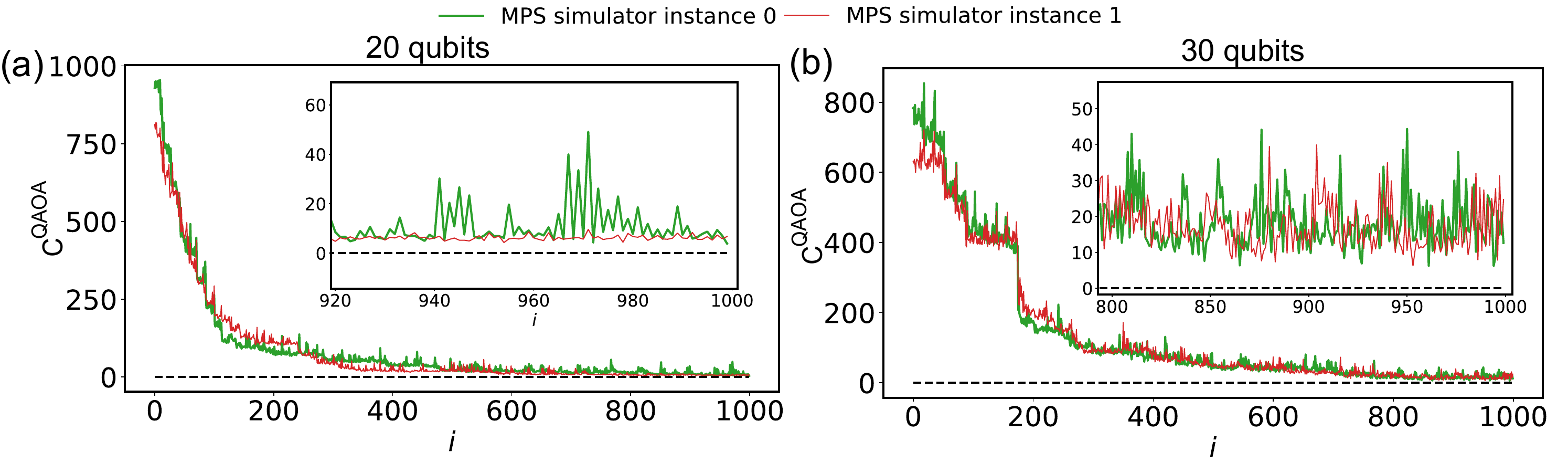}
    \caption{Illustration of the scalability of Quantum Approximate Optimization Algorithm (QAOA) to larger-scale simulations. We evaluate the cost function [Eq.~\ref{cost2}] for 20- and 30-qubit problems, as shown in panels (a) and (b) respectively, using a noiseless matrix product state (MPS) simulator, where $i$ denotes the number of optimization iterations. The zoomed-in regions highlight growing instability in the optimization landscape as the system size increases.   Despite notable instability at these scales, the cost function ultimately converges to a reasonable energy close to the exact ground-state energy, orders of magnitude lower than that of the initial state.}
    \label{fig:mps}
\end{figure*}

\section{Benchmarking Results}\label{sec3}
\subsection{Benchmarking simulations of QAOA}
In this section, we first evaluate the performance of the QAOA method, with the structure of the variational circuits detailed in FIG.~\ref{fig:circuit2}.  During the optimization process, the cost function, defined by the energy difference in Eq.~\eqref{cost}, is minimized to approximate the ground state of the target Hamiltonian.

\subsubsection{Noiseless QAOA simulations}

As depicted in FIG.\ref{fig:qaoa} (a), we implement the QAOA procedure using a noiseless local simulator. \rz{Here, in each optimization iteration, we perform a full round of circuit simulation (distinct from an actual circuit run in Qiskit) and subsequently update the trainable parameters based on the feedback.} For these noiseless simulations, the optimization exhibits near-perfect convergence towards $E_g$ after approximately $600$-$1000$ iterations. However, this high accuracy comes at the cost of significant quantum resources, requiring over $600$ circuit evaluations on quantum hardware. 
The lowest discrepancy ${\rm Min}|E_{g}-E_i|$ from the true ground state energy $E_g$ achieved throughout the QAOA optimization process in each instance are presented in FIG.~\ref{fig:qaoa} (b). Here, the optimized circuits converge to $C^{{\text{QAOA}}}(\boldsymbol{s})<2.5$, indicating good approximations to the ground state, as further detailed in Appendix~\ref{subqaoa}. \rz{Note that different problem instances correspond to Hamiltonians with distinct parameter sets derived from independently generated asset data, as detailed in Appendix~\ref{data}. Consequently, their ground states are associated with different optimal bitstrings.}
That said, even in the absence of noise, the finite number of circuit executions introduces intrinsic sampling errors, preventing convergence to the exact ground state.

\subsubsection{QAOA simulations with controlled noise}

We then conduct controlled noisy simulations of QAOA on a local simulator, and the  {effect of CX gate noise on the trajectory} of this optimization process is presented in FIG.~\ref{fig:qaoa} (c). Under increasing CX noise ({error rates}), the optimization becomes highly unstable and exhibits no clear convergence, highlighting the significant impact of noise on QAOA performance. This trend is further illustrated in Fig.\ref{fig:qaoa}(d), where it is evident that the minimum energy achieved during optimization increases rapidly with growing CX gate error rates.  {Even for a $0.001$ CX error rate, which is comparable to that in contemporary IBM quantum processors, $E_g-E_i$ is already far larger than that from noiseless simulations.}

%\CH{CH: seems that even 0.001 CX errors give rise to far larger $E_i$ than in the noiseless simulations. For what value of CX error rate does the $E_i$ approach that of the noiseless simulations?} \rz{the reason is that noiseless (MPS-based) and noisy (sampling-based) simulators are very different.}

\begin{figure*}
	\centering
	\includegraphics[width=.8\linewidth]{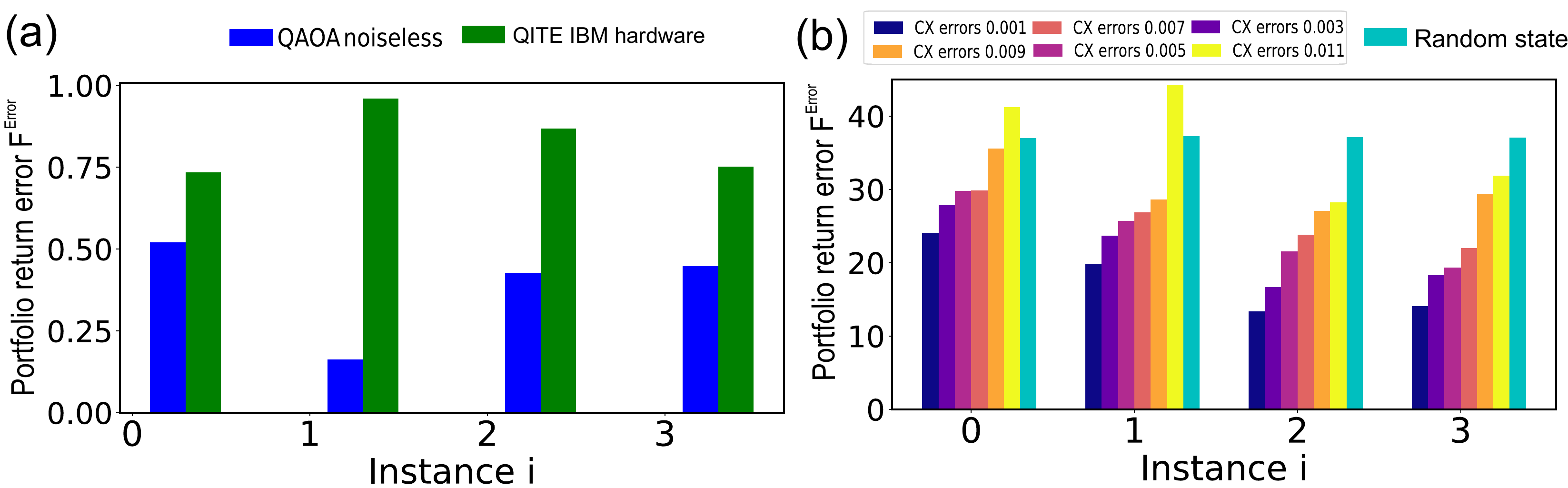}
	\caption{Assessment of QAOA and QITE performance via return errors $F^{\rm Error}=F^{\rm ideal}-F^{\rm circuit}$ [Eq.\ref{error}]. (a)  Return errors for four selected instances obtained from noiseless QAOA simulations (blue) and QITE executed on IBM Quantum hardware (green) {with typical 2-qubit gate error rate of $0.007$}. Both methods yield results that closely approximate the exact returns, with QAOA achieving slightly better accuracy with the benefit of ideal noiseless conditions. (b)  Return errors from noisy QAOA simulations on a local simulator with a range of two-qubit (CX) gate error rates representative of that in cutting-edge digital quantum processors. Higher noise levels lead to a increase in errors across all instances, highlighting the sensitivity of QAOA to noise.  Cyan bars denote outcomes computed under random states, which represent return errors for completely unoptimized reference cases. %exhibiting substantially large errors. %CH: not true
	Error bars represent standard deviations computed over 10 random samples. The selected instances are the same as those used in FIG.~\ref{fig:qaoa}.} 
	\label{fig:qaoaqite}
\end{figure*}

In this work, QAOA optimization is performed using the COBYLA optimizer \cite{polak2012optimization}, while the corresponding quantum state measurements are collected from a noisy simulator. While COBYLA is designed to handle constrained problems efficiently, it relies on iterative evaluations of the cost function. When those evaluations are affected by hardware noise, the optimizer struggles to navigate the resulting rugged energy landscape, often leading to poor convergence behavior.

\subsubsection{QAOA simulations -- scaling to larger systems}

To further evaluate the scalability of QAOA,  we simulate its performance on 20- and 30-qubit systems using a matrix product state (MPS) simulator. Here, we present the results of two instances in FIG.~\ref{fig:mps}. For 20-qubit instances, the QAOA optimization process exhibits relatively stable convergence, and the optimal outcome $|E_{g}-E_{i}|\approx5$ is very small compared to the overall energy scale. Despite notable fluctuations during the optimization process of the 30-qubit instances, the final stage can reach a similar energy level of $|E_{g}-E_{i}|\approx 10$,  {comparable} with the results in smaller systems. These results demonstrate that QAOA can effectively approach optimal solutions even at intermediate system sizes. Since poor convergence under noise has already been observed in smaller systems, and noisy simulations at the 20-qubit scale lead to substantial computational cost, we do not present noisy simulations for these larger instances.

\subsection {Comparison between QAOA and QITE}

\subsubsection{Benchmarking from predicted returns}

In this section, we compare the performance of QAOA and QITE in terms of their predicted returns $F$, as defined and computed by Eq. \ref{opt}. Using the same representative instances, we compute the return error between the circuit outputs and the exact {(ideal)} results, defined as
\begin{equation}\label{error}
F^{\rm Error}=F^{\rm ideal}-F^{\rm circuit},
\end{equation} 
where $F^{\rm ideal}\approx 0$, and $F^{\rm circuit}<0$ denotes loss in portfolio value due to {suboptimal} outcomes. FIG.~\ref{fig:qaoaqite} illustrates the simulated results of such returns. FIG.~\ref{fig:qaoaqite} (a) demonstrates that, remarkably, the QITE algorithm executed on IBM quantum hardware ({with $\approx 0.007$ ECR/CX error rate, as obtained from Fig. S1 from Appendix.~\ref{ibm}}) predicts return values that 
exhibit very low $F^{\rm Error}< 1$, almost comparable to that from the QAOA instances obtained in ideal (noiseless) settings. Note that because the implementation of QAOA requires a large number of circuit runs, so to save cost on IBM Quantum hardware, we only conduct QAOA simulations on local simulators which can be calibrated to realistic noise levels of actual quantum hardware (see Appendix.~\ref{ibm}).

To put the excellent performance of the QITE hardware runs into perspective, we compare the return error $F^{\rm Error}$ from QAOA with simulated noise of similar magnitude [FIG.~\ref{fig:qaoaqite} (b)]. Saliently, even the smallest levels of noise significantly affect QAOA performance, with return errors exceeding $F^{\rm Error}>20$, and increasing further as noise levels rise. Notably, the optimal results obtained from QAOA are not stable: under a moderately elevated noise level ($0.011$), the first two instances exhibit a sharp increase in return error. This instability arises because noise can introduce temporal fluctuations in the measured cost function, leading to a non-smooth landscape in optimization. As a result, the performance of QAOA becomes highly sensitive to both the specific problem instance and the stochastic behavior of the circuit runs. 

To further contextualize our $F^{\rm Error}$ results, we also compute returns from reference states with uncorrelated uniformly random coefficients, which represent completely unoptimized outcomes~\footnote{These random states are not produced by random circuits; instead, we generate them numerically as $\ket{\psi}=\sum_{i}a_{i}\ket{\psi_{i}}/||\sum_{i}a_{i}\ket{\psi_{i}}||$, which are normalized. Here, the coefficients $a_{i}$ are randomly sampled from a uniform distribution within $[0,1]$, and $\ket{\psi_{i}}$ range over the computational basis states.}.
As shown in FIG.\ref{fig:qaoaqite}(b) (cyan bars), these unoptimized reference random states exhibit large errors of $F^{\rm Error}\approx 38$, which substantially exceed those in FIG.\ref{fig:qaoaqite}(a), and are also larger than the low-noise QAOA results also plotted in FIG.~\ref{fig:qaoaqite}(b). However, under significant but realistic noise levels, such as a CX error rate of $0.011$, QAOA performance error approaches the level of these fully unoptimized outcomes, offering no optimization value at all.

\begin{figure*}
	\centering
	\includegraphics[width=0.9\linewidth]{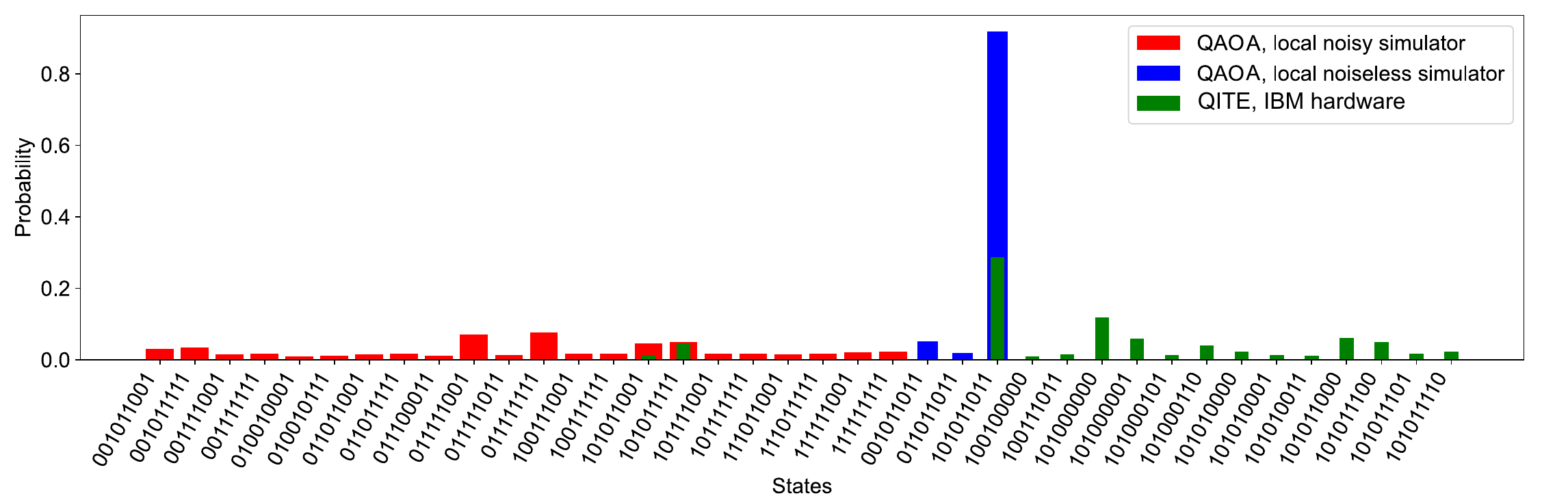}
	\caption{Probability distribution of the optimal states obtained from QAOA and QITE for the representative instance 0 (FIG.~\ref{fig:qaoa}), so as to compare their convergence towards the ideal bitstring 101011011 (instance 0 in FIG.~\ref{fig:qaoa}). Results are shown in green, blue, and red, corresponding to IBM Quantum simulations of the QITE method, noiseless QAOA simulations, and noisy QAOA simulations (local simulator with realistic noise levels of $0.007$ (two-qubit gates), Appendix~\ref{ibm}), respectively. In simulations on IBM hardware, prominent peaks (green) successfully identify the optimal state. For the noiseless QAOA simulations (blue), the distribution exhibits an even more distinct sharp peak at the optimal solution bitstring. In the noisy QAOA simulation (red), noise effects significantly impact performance, preventing the identification of the optimal state. 
    }
	\label{fig:ibmqaoa}
\end{figure*}

\subsubsection{Benchmarking from measured optimal state distribution}

We now examine the actual solution states obtained from both QITE and QAOA, which represents the ideal portfolio composition. In principle, the ideal optimal solution corresponds to a single bitstring with the highest return. Thus, to assess how the outputs of QAOA and QITE deviate from the ideal solution, we examine a representative problem instance and analyze the full probability distributions of measured quantum states. 

As shown in FIG.~\ref{fig:ibmqaoa}, the QITE results (green bars) obtained from simulations on IBM hardware exhibit a  {reasonably} prominent peak at the bitstring 101011011 corresponding to the exact optimal solution. Although the obtained state probability distribution is broadened due to the presence of noise, the optimal bitstring remains clearly identifiable. For comparison, the blue bars in FIG.~\ref{fig:ibmqaoa} display the probability distribution obtained from the noiseless QAOA simulation with the same instance. In this ideal noiseless setting, QAOA yields a nearly 
perfectly localized state on this optimal bitstring (see FIG.~\ref{subqaoa}) 
In contrast, QAOA results (red bars) conducted on a local simulator under realistic errors (see Appendix.~\ref{ibm}), fail to reveal any distinct peak corresponding to the optimal solution. This behavior highlights the instability under noise, consistent with the unstable convergence trends observed in FIG.~\ref{fig:qaoa}.

\begin{figure*}
    \centering
    \includegraphics[width=1\linewidth]{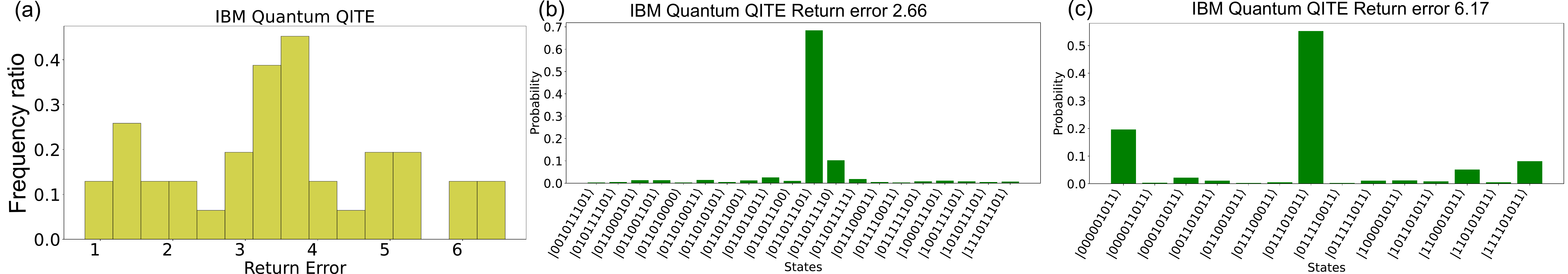}
    \caption{Further performance assessment of the Quantum Imaginary-Time Evolution (QITE) algorithm in noisy quantum processors. We perform additional simulations for $50$ problem instances to evaluate the robustness of QITE in small-scale ($9$ qubits) settings. (a) Yellow bars denote the histogram of return errors $F^\text{Error}$ [Eq.\ref{error}] for simulations conducted on the IBM Quantum processor, spanning relatively small values of $1~to~6$. (b) and (c) State {probability} distributions for two selected instances from (a), with corresponding return errors of $2.66$ and $6.17$, respectively. In both cases, the optimal strings remain clearly identifiable despite the non-negligible return errors. The Hamiltonian corresponds to a system size of $L=9$ {qubits}. \rz{Note that these 50 instances represent different Hamiltonians, and more details are shown in Appendix.~\ref{data}.}}
    \label{fig:return}
\end{figure*}

\begin{table*}
\centering
\begin{tabular}{|p{5.5cm}|p{6.2cm}|p{5.4cm}|}
\hline
\textbf{Methods} & \textbf{QAOA} & \textbf{QITE} \\
\hline
\textbf{Noiseless Performance (FIG.~\ref{fig:qaoaqite})} & Excellent convergence to exact ground states & Slight discrepancies due to imperfect circuit optimization \\
\hline
\textbf{Noisy Performance (FIG.~\ref{fig:ibmqaoa})}
\qquad\qquad\qquad Two-qubit gate error rate $\sim 10^{-2}$ & Unstable; fails to converge reliably & Robust; optimal state still identifiable \\
\hline
\textbf{Resource Requirements} & High quantum resources (extensive circuit evaluations); light classical cost & Moderate quantum resources; heavy classical optimization required  (Appendix.~\ref{QITE}) \\
\hline
\textbf{Scalability on NISQ Devices (FIG.~\ref{fig:mps})} & Better scalability if noise is mitigated & Limited by classical optimization cost \\
\hline
\end{tabular}
\caption{Comparison between the two approaches: QAOA and QITE for portfolio optimization {and other problems based on finding Ising ground states}.}
\label{tab:qaoa_vs_qite}
\end{table*}

To further validate the robustness of QITE, we simulate the QITE circuit for additional 50 portfolio optimization instances {on the IBM quantum computer, also with $L=9$ qubits}.
These simulations are performed on both IBM Quantum hardware. As shown in FIG.~\ref{fig:return}, the return errors $F^\text{Error}$ range from about 1 to 6, which is much smaller than those from QAOA with similar simulated CX gate error rate of $0.007$ ({see Appendix.~\ref{ibm}}), as previously shown in FIG.~\ref{fig:qaoaqite}.
 \rz{Note that the return error range can arise from variations in the optimization quality of circuits across different instances, as well as from noise fluctuations in simulations.} FIGs.\ref{fig:return}(b,c) present the measured solution state distributions for two representative instances drawn from FIG.\ref{fig:return}(a). Despite higher return errors of $2.66$ and $6.17$, the optimal bitstring still remain prominent.

\section{Conclusion}
In this work, we benchmarked two quantum algorithms, QAOA and QITE, which are widely used in digital quantum simulations for solving {not just the Markowitz portfolio optimization problem, but also other diverse problems with quantum Ising formulations} \cite{lucas2014ising,kalinin2022computational,cipra2000ising,kalinin2020complexity}. Our results show that the QAOA method achieves excellent convergence to the exact ground state in noiseless simulations when hardware noise is negligible or {extensively}  mitigated. However, the QAOA method is substantially sensitive to noise, particularly due to its reliance on deep parameterized circuits and repeated quantum evaluations, which are highly sensitive to gate and readout errors. {In contrast, the QITE method exhibits much stronger robustness to realistic noise levels, achieving comparably optimal solutions on actual quantum hardware runs as noiseless QAOA simulated runs.}

From a resource perspective, as touched on earlier in Section.~\ref{vqe},  QAOA demands extensive quantum resources, typically requiring hundreds to thousands of circuit executions, but relatively low classical computational cost. However, QITE, on the other hand, requires less quantum cost but substantial classical computational cost due to the need for extensive circuit training. As illustrated in FIG. \ref{fig:qaoa} and FIG.~\ref{fig:mps}, the 9-qubit instance of QAOA requires %approximately 400 
a few hundred iterations for reasonable convergence, which increases to around 1000 iterations for the 20-qubit case.
For QITE, only a single round of circuit simulation is required, equivalent to the cost of one QAOA iteration, resulting in markedly lower quantum resource requirements. For problems of the same scale, QAOA requires hundreds of times more quantum resources than QITE.

On quantum hardware, qubit measurements are straightforward and can be performed on systems with over 100 qubits. As a result, in terms of scalability, QAOA holds greater theoretical promise for tackling larger problem instances, if hardware noise can be effectively mitigated. A detailed side-by-side comparison of both methods is summarized in Table~\ref{tab:qaoa_vs_qite}.

Overall, our comparative analysis of the noise resilience and scalability in quantum optimization strategies suggests that the choice between QAOA and QITE should be guided by the specific noise characteristics of the quantum device, the availability of classical computational resources, and the complexity and scale of the optimization task \cite{ajagekar2019quantum,de2021materials,marzec2016portfolio,buonaiuto2023best,carvalho2021error,shen2023proposal,kottmann2021quantum}.

\section{Acknowledgements} 
We acknowledge the use of IBM Quantum services for this work. The views expressed are those of the authors and do not reflect the official policy or position of IBM or the IBM Quantum team. All data and codes of this work are available from the corresponding authors upon reasonable request. This work is supported by the Singapore Ministry of Education Academic Research Fund Tier-I preparatory grant (WBS no. A-8002656-00-00) and  MOE's Tier-II grant (WBS no: A-8003505-00-00).

%%%%%%%%%%%%%%%%%%%%%%  References %%%%%%%%%%%%%%%%%%%%%%
%\bibliographystyle{apsrev4-1}
%\bibliographystyle{plain}
%\bibliography{references}
\bibliography{ref}

\onecolumngrid
\flushbottom
\newpage
\appendix
\setcounter{equation}{0}
\setcounter{figure}{0}
\setcounter{table}{0}
\setcounter{section}{0}
\renewcommand{\theequation}{S\arabic{equation}}
\renewcommand{\thefigure}{S\arabic{figure}}
\renewcommand{\thesection}{S\arabic{section}}
\renewcommand{\thepage}{S\arabic{page}}
\renewcommand{\thetable}{S\arabic{table}}
%
%\newpage
%\section*{Supplementary informations}
\section{Correspondence between the  Markowitz Portfolio model and the Ising Hamiltonian}\label{ap1}
\hzc{
In this section, we detail the construction of the Ising Hamiltonian for the Markowitz portfolio selection problem described by Eq.~\ref{opt}. We begin by reformulating the total expected return using binary variables, a necessary step for quantum computing applications \cite{grant2021benchmarking}.}

\hzc{We use $z_u$ to represent the investment in asset $u$, quantified as an integer number of quantized minimum investment fractions. We can express each integer $z_u$ using a standard binary representation with $w$ binary variables, $x_{i(u,k)} \in \{0,1\}$:
\begin{equation}
z_u = \sum_{k=1}^w 2^{k-1} x_{i(u,k)},
\label{eq:supp-binary-rep}
\end{equation}
where $i(u,k)=(u-1)w+k, k=1,2,...,w.$
In this expression, the variables $\{x_{i(u,1)}, \dots, x_{i(u,w)}\}$ are the binary digits of $z_u$. For example, if we use $w=3$ bits to represent the investment in the $u-$th asset, an amount of $z_u=5$ (which is $101_2$ in binary) would be encoded by setting $x_{i(u,1)}=1$, $x_{i(u,2)}=0$, and $x_{i(u,3)}=1$, since $5 = (1 \cdot 2^0) + (0 \cdot 2^1) + (1 \cdot 2^2)$.
Substituting this binary expansion of $z_u$ [Eq.~\ref{eq:supp-binary-rep}] into the formula for total expected return in Eq.~\ref{opt} yields :
\begin{equation}
	\sum_{u=1}^m r_u z_u = \sum_{u=1}^m r_u \left( \sum_{k=1}^w 2^{k-1} x_{i(u,k)} \right) = \sum_{u=1}^m \sum_{k=1}^w 2^{k-1} r_u x_{i(u,k)}.
\end{equation}
Here, $m$ is the total number of assets, $w$ is the number of binary variables (or ''qubits'') allocated to each asset, and $r_u$ is the expected return for the $u$-th asset. The function $i(u,k) = (u-1)w+k$ simply assigns a unique index to each binary variable across all assets.}

The corresponding covariance term is rewritten as follows
\begin{equation}
	-\sum_{u,v}^m c_{u,v} z_u z_v = -\sum_{u,v}^m \sum_{k=1}^w \sum_{k'=1}^w 2^{k-1} 2^{k'-1} x_{i(u,k)} x_{j(v,k')}. 
\end{equation}
The allocation constraint is given by:
\begin{equation}
	-(\sum_{u=1}^m p_w b z_u-b)^2  = -(\sum_{u=1}^m \sum_{k=1}^w 2^{k-1} p_wbx_{i(u,k)} -b)^2,
\end{equation}
where $b$ is the budget and $p_w = 1/2^{w-1}$ is the quantized minimum investment fraction.

Based on such a binary representation, Eq.~\ref{opt} from the main text can be rewritten as:
\begin{equation}
	\mathop{max}\limits_{z} [\theta_1 \sum_{i=1}^n r_i x_i -\theta_2 \sum_{i,j=1}^n c_{i,j} x_{i}x_{j}- \theta_3(\sum_{i=1}^n 2^{k-1} b p_w x_i-b)^2],
\end{equation} with the following rescaling parameters:
\begin{equation}
	\begin{aligned}
		n&= mw, \\
		r_i&= 2^{k-1}r_u, \\
		c_{i,j} &= 2^{k-1} 2^{k'-1} c_{u,v}.
	\end{aligned}
	\label{rescale_params}
\end{equation}
The above optimization is equivalent to a Quadratic Unconstrained Binary Optimization (QUBO) problem, expressed as:
\begin{equation}
	\mathop{min}\limits_x \Big( \sum_i^n q_ix_i + \sum_{i,j}^n Q_{i,j}x_i x_j + \gamma \Big),
\end{equation}
where the parameters are defined as:
\begin{equation}\label{rescale}
	\begin{aligned}
		q_i &= -\theta_1r_i -2 \theta_3 b^2p_w 2^{k-1},\\
		Q_{i,j} &= \theta_2 c_{i,j}+ \theta_3 b^2 p_w^2 2^{k-1} 2^{k'-1},\\
		\gamma &= \theta_3 b^2.
	\end{aligned}
\end{equation}

By transforming $x_i = \frac{1}{2} (1+\sigma_i^z)$, the above optimization can be reduced to finding the ground state of the following Ising Hamiltonian:
\begin{equation}\label{supph}
	H= \sum_i h_i \sigma_i^z + \sum_{i,j} J_{i,j}\sigma_i^z \sigma_j^z + \delta,
\end{equation}
with the corresponding parameters:
\begin{equation}
	\begin{aligned}
		J_{i,j} &= \frac{1}{4} Q_{i,j}, \\
		h_i &= \frac{1}{2}q_i + \frac{1}{2} \sum_j Q_{i,j}, \\
		\delta &= \frac{1}{4} \sum_{i,j} Q_{i,j} + \frac{1}{2}\sum_i^n q_i + \gamma.
	\end{aligned}
	\label{eq:ising model params}
\end{equation}
In our implementation, we discard the constant term $\delta$, as it does not affect the optimization process.

\section{Dataset Preparation
}\label{data}
In this section, we outline the method used to prepare our dataset, which simulates the historical prices for financial assets. The asset price dataset is structured as a matrix of dimensions $(N_f, m)$, where $N_f$ represents the number of historical price points per asset, and $m$ is the number of assets under consideration. Each column in this matrix corresponds to the price history of a distinct asset.

The dataset preparation process begins by setting the \hzc{price generator $a_{u,0}$}, for the $u$-th asset.  This price generator is randomly selected from the range [$b/10, b$], where $b$ is the total budget. To generate brownian-like temporal price fluctuations, each subsequent historical price point, $a_{u,l}$, is then generated within a fluctuation range of $75\%$ to $125\%$ of the price generator, specifically, $a_{u,l}= (1+\alpha) a_{u,0}$. This procedure generates the $u$-th column of the price data matrix. To construct the full matrix, this process is repeated across all assets. Based on the generated price data, the expected return $r_u$ and covariance matrix $c_{u,v}$ are calculated using Eq.~\ref{eq:r_u}. and Eq.~\ref{eq:c_uv} respectively.

With the calculated price data, we proceed to generate the Ising Hamiltonian problem instances. \hzc{This is achieved by selecting one combination of the parameters $\theta_{1}, \theta_2, \theta_3$ subject to the constraint $\sum_i \theta_i = 1$, where different choices of $\bm{\theta}$ correspond to distinct investment preferences.} These parameters, which represent different investor preferences, are integrated with the calculated $r_u$ and $c_{u,v}$. Subsequently, the rescaling process described in Eq.~\ref{rescale_params} is applied. A detailed workflow (9-qubit instance in the main text) is demonstrated as follows, and the corresponding code and data are available at \cite{da}:

\begin{enumerate}
\item \textbf{Input parameters:} The number of assets is \( m = 3 \), with each asset having \( N_f = 100 \) price points. Each asset is represented using \( w = 3 \) binary bits (slices), with $p_w = 1/ 2^{w-1}$. A total of 100 problem instances are generated, with a total budget \( b = 10 \). Additional parameters are set to \(\theta_1 = 0.8\), \(\theta_2 = 0.1\), and \(\theta_3 = 0.1\).

\item \textbf{Price generation:} For each asset \( u \), randomly initialize the price generator \( a_{u,0} \) within the range \([b/10, b]\). Generate historical prices using \( a_{u,l} = (1 + \alpha) a_{u,0} \),  where \( \alpha \) is a random variable uniformly sampled from \([-0.25, 0.25]\).

\item \textbf{Covariance matrix and expected return:} For each instance:
\begin{itemize}
    \item Compute the covariance matrix \( c_{uv} \) as defined in Eq.~3 of the main text:\\
	\( c_{u,v} = \frac{b^2p_w^2}{a_{u,N_f}a_{v,N_f}} \cdot \frac{ \sum_{l=1}^{N_f} (a_{u,l}-\bar{a}_u) (a_{v,l}-\bar{a}_v)}{N_f -1}, 
\)
where $\bar{a}$ denotes the mean price of  the assets and $p_w=2^{1-w}$.
    \item Compute daily returns \( \Delta a_{u,l} = a_{u,l+1} - a_{u,l} \) and the expected return \( r_u \) from Eq.~2 of the main text:
\(	r_u = \frac{b p_w}{a_{u,N_f}}\cdot \sum_{l=1}^{N_f} \frac{a_{u,l+1}-a_{u,l}}{N_f-1}.\)
\end{itemize}
\item \textbf{Rescaling:} Rescale \( c_{uv} \) and \( r_u \) to $Q_{i,j}$ and $q_i$ (parameters for QUBO problems) according to Eq.~\ref{rescale_params} and Eq.~\ref{rescale}. 
\item \textbf{Ising coupling and field parameters:} The total number of binary variables is \( m \cdot w = 9 \). For binary indices \( i, j \in \{ 1, \dots, 9 \} \), the corresponding Ising coupling \( J_{ij} \) and Ising field \( h_i \) are computed based on Eq.~\ref{eq:ising model params}: $J_{i,j} = \frac{1}{4} Q_{i,j}, 
		h_i = \frac{1}{2}q_i + \frac{1}{2} \sum_j Q_{i,j}$.
\end{enumerate}

\rz{In this work, for the 9-qubit problem, we generate 100 distinct instances, each associated with a specific Hamiltonian. Consequently, the optimization landscape varies across instances, and the optimal solution bitstring is not universally fixed.}

\section{IBM quantum device used}\label{ibm}
\begin{figure}[h]
	\centering
	\includegraphics[width=0.9\linewidth]{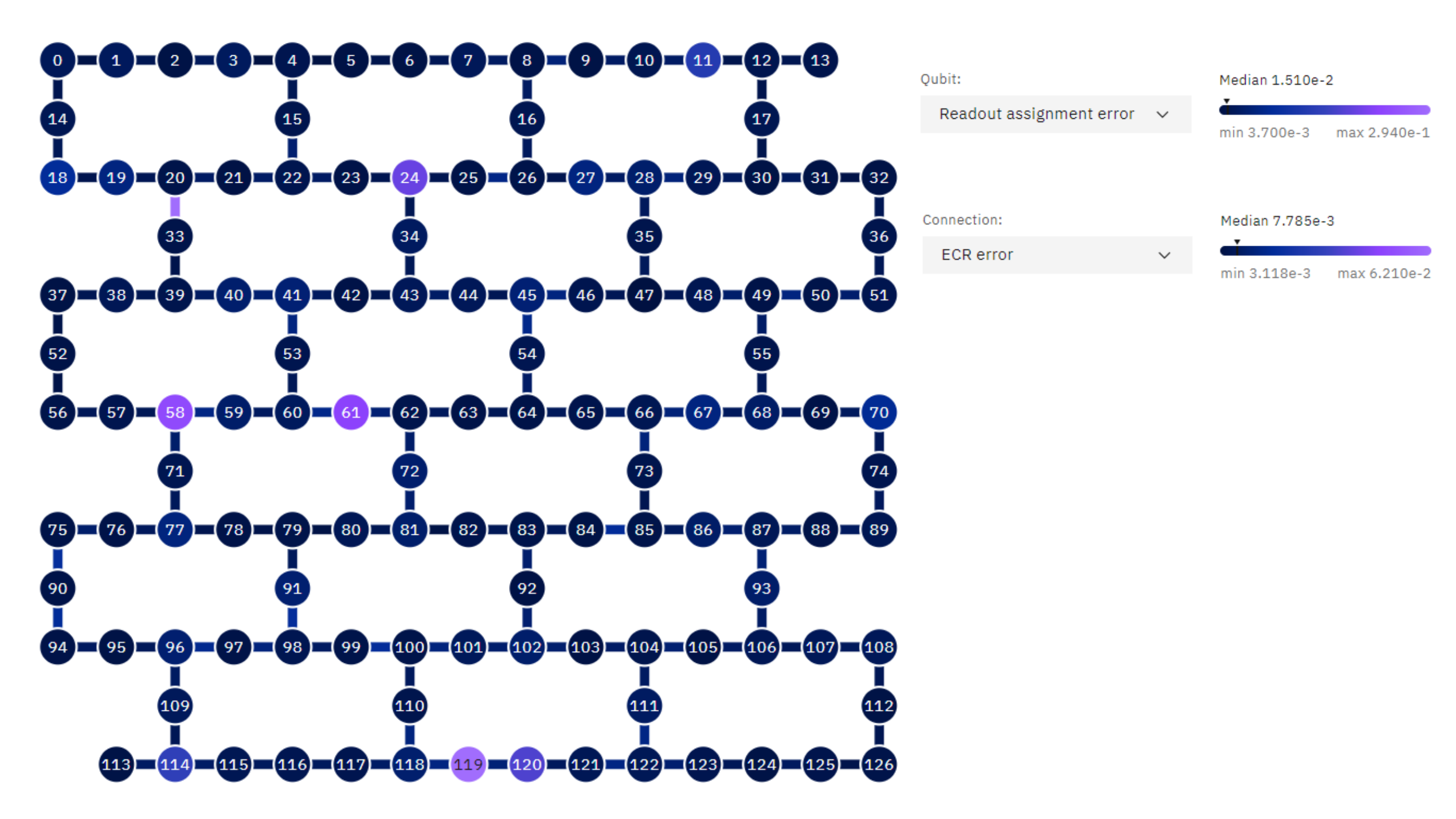}
	\caption{The error map of the IBM "Brisbane" device. For simulations in our work, we select the qubit chain $[0,1,2,3,4,5,6,7,8,9,10,11,12]$, with the mean ECR error rate around $0.007$.}
	\label{fig:errormap}
\end{figure}
For simulations in this work, we implement jobs on the IBM "Brisbane" device, and details of the gate errors are shown in FIG.~\ref{fig:errormap}. Here, we select the qubit chain $[0,1,2,3,4,5,6,7,8,9,10,11,12]$ to be used for our computations. Note that simulations shown in  FIG.~\ref{fig:ibmqaoa} are based on a local noise model calibrated using the error parameters extracted from the ``Brisbane" device \cite{qiskit2024}.

\section{Quantum imaginary-time evolution}\label{QITE}
In this section, we describe our approach for simulating the non-unitary evolution operator $U_{\rm non}=e^{-\beta H}$. Since quantum circuits are inherently unitary, this non-unitary operation must be embedded into a larger unitary operator that can be physically implemented. This embedding can be constructed in the following form:
\begin{equation}\label{nu}
	U=\left[\begin{array}{cc}
		uU_{\rm non} & B \\                                                                       
		C & D
	\end{array}\right].
\end{equation}
where the scaling factor $u^{-2}$ denotes the maximum eigenvalue of $U^{\dagger}_{\rm non}U_{\rm non}$
and the lower-left block $C$ is numerically solved, $C=A\sqrt{I-u^{2}\Sigma^{2}}E^{\dagger}$ with identity matrix $I$. The matrix elements $A$, $E$, and $\Sigma$ are obtained from the singular value decomposition (SVD) of the operator $U_{\rm non}$: $U_{\rm non}=A\Sigma E^{\dagger}$
\cite{shen2025observation,lin2021real}, where $\Sigma$ is a diagonal matrix containing the singular values, and $A$, $E$ are unitary matrices.

To construct the full unitary operator $U$, we first perform QR decomposition on an associated ansatz matrix:
\begin{equation}\label{supussh}
	U^{\prime}=\left[\begin{array}{cc}
		uU_{\rm non} & I\\                            
		C& I
	\end{array}\right],
\end{equation}
where the right-side blocks are simply taken to be identity matrices. To recover $U$, we can apply the QR decomposition to $U^{\prime}$, such as to obtain
\begin{equation}
	U^{\prime}=UM,
\end{equation}
where $U$ ($UU^{\dagger}=I$)  is the desired unitary operator, and $M$ is an upper triangular matrix. 

Here, we present a concrete example to demonstrate how $U_\text{non}$ can be executed from the $U$ circuit.  We can consider an initial composite state $\ket{\psi}\otimes\ket{0}$, where \(\ket{\psi}\) is the input state for physical problems, and \(\ket{0}\) is for the ancilla qubit. The action of the  unitary operator \(U\) on this state can be expressed as
\begin{equation}
	U \big( \ket{\psi} \otimes \ket{0} \big) 
	= \big( u\,U_{\mathrm{non}} \ket{\psi} \big) \otimes \ket{0} 
	+ \big( C \ket{\psi} \big) \otimes \ket{1},
\end{equation}
where \(U_{\mathrm{non}}\) denotes the desired non-unitary evolution. Then, we perform a projective measurement on the ancilla qubit. The post-selecting outcome with \(\ket{0}\) is recorded, and the undesired branch associated with \(\ket{1}\) is discarded. The resulting normalized state of system qubits is 
\begin{equation}
	\frac{U_{\mathrm{non}} \ket{\psi}}{\| U_{\mathrm{non}} \ket{\psi} \|}.
\end{equation}

{The implementation of this unitary matrix $U$ can be achieved through variational circuits. 
Note that the Hamiltonian defined by Eq.~\ref{supph} includes long-range couplings, so the optimization of variational circuits requires deep layers.  As a result, the QITE method in this work is primarily applied to small-scale systems, where circuit depth and computational cost remain manageable. For example, the 9-qubit instance presented in the main text requires only 2 layers in the QAOA circuit, whereas the corresponding QITE circuit demands between 4 and 8 layers to achieve convergence. Despite the scaling challenges, larger instances, such as those with 20 and 30 qubits, can still be addressed using shallow QAOA circuits with just 2 layers.}

{For the optimization of QITE circuits,  we set a convergence threshold of $C^\text{QITE}<0.1$ [Eq.~\ref{E}]. This ensures that the evolved state closely approximates the ground state, and also keeps the circuit depth moderate and computationally manageable.} \rz{As shown in FIG.~7(a), such imperfect convergence can give rise to a broad spread in the return error.}

\section{Quantum annealing and adiabatic evolution}
Quantum annealing is another approach designed to solve optimization and minimization problems \cite{yarkoni2022quantum,hauke2020perspectives,santoro2006optimization}. This method is based on the adiabatic evolution of a quantum system governed by a time-dependent Hamiltonian that interpolates between an initial Hamiltonian $H_i$ and the target Hamiltonian $H_p$. This evolution is described by: $H(\lambda(t)) = (1-\lambda(t))H_{i} + \lambda(t) H_{p},$ where $\lambda(t)$ is a time-dependent annealing parameter satisfying the boundary conditions $\lambda(0)=0, \lambda(T)=1$.  The system is initially prepared in the ground state of $H_i$ and evolves under $H_{ad}(\lambda(t))$, with the target Hamiltonian gradually turned on. According to the quantum adiabatic theorem, if the evolution proceeds slowly enough, the system remains in its instantaneous ground state and eventually transitions to the ground state of $H_{p}$.

In general, quantum annealing can be naturally implemented on analog quantum simulators, where the evolution on such quantum platforms closely follows a continuous-time annealing schedule. However, the simulator employed in this work is digital, and emulating quantum annealing dynamics on such a platform requires extremely deep quantum circuits to Trotterize the continuous evolution.  These deep circuits are highly susceptible to noise and decoherence. Given this, we do not include quantum annealing in our performance benchmarks.

\section{Additional results}\label{subqaoa}
\begin{figure}[h]
    \centering
    \includegraphics[width=0.75\linewidth]{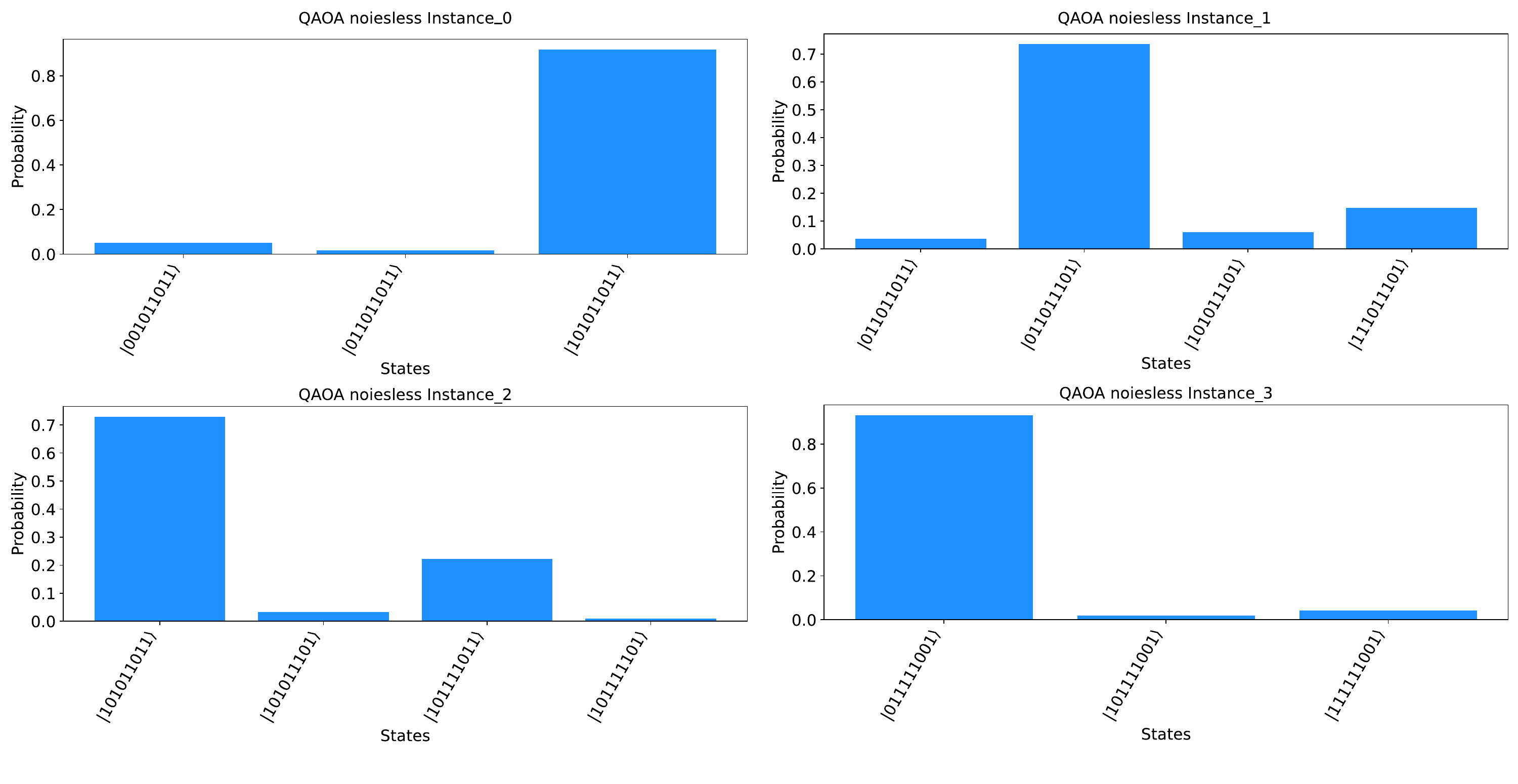}
    \caption{Probability distributions from noiseless QAOA simulations for four representative problem instances (FIG.~\ref{fig:qaoaqite}, main text). In all cases, a distinct peak appears at the optimal bitstring, indicating successful convergence.}
    \label{fig:qaoa_data}
\end{figure}
Here, we present additional results to demonstrate that the noiseless optimization of QAOA can reach reasonable convergence. These results are presented in FIG.~\ref{fig:qaoa_data}, for the four instances shown in FIG.~\ref{fig:qaoaqite} (main text). These results clearly demonstrate that, in the absence of noise, QAOA can converge toward the ground state, successfully identifying the optimal bitstring for each instance. \rz{Note that for different instances, the corresponding Hamiltonian might also contain different parameters corresponding to different instances of generated asset data, leading to the convergence to distinctive bitstrings.}

\flushbottom
%%%%%%%%%%%%%%%%%%%%%%  References %%%%%%%%%%%%%%%%%%%%%%
%\bibliographystyle{apsrev4-1}
%\bibliographystyle{plain}
%\bibliography{references}
%\subsection*{\normalsize Supplementary references}
%\bibliography{ref}

\end{document}